\documentclass[12pt]{article}
\usepackage{epsfig}

\def\dfrac#1#2{{\displaystyle {#1 \over #2}}}

\def\dint{\displaystyle \int }

\def\simge{\mathrel{\rlap{\raise 0.511ex \hbox{$>$}}{\lower 0.511ex 
 \hbox{$\sim$}}}}
\def\simle{\mathrel{\rlap{\raise 0.511ex \hbox{$<$}}{\lower 0.511ex 
 \hbox{$\sim$}}}} 
\def\slash#1{\setbox0=\hbox{$#1$}\dimen0=\wd0 \setbox1=\hbox{/} \dimen1=\wd1 
 \ifdim\dimen0>\dimen1 \rlap{\hbox to \dimen0{\hfil/\hfil}} #1 
 \else \rlap{\hbox to \dimen1{\hfil$#1$\hfil}} / \fi}                   

\newcommand{\be}{\begin{equation}}
\newcommand{\ee}{\end{equation}}
\newcommand{\bea}{\begin{eqnarray}}
\newcommand{\eea}{\end{eqnarray}}
\newcommand{\nn}{\nonumber}
\newcommand{\Lam}{\Lambda_{\rm QCD}}
\newcommand{\ms}{{\overline{\rm{MS}}}}
\newcommand{\as}{\alpha_{ s}}

\newcommand{\aspu}{\dfrac{\alpha_{s}(1/x)}{4\pi}}
\newcommand{\mev}{\ {\rm MeV}}
\newcommand{\gev}{\ {\rm GeV}}
\newcommand{\g}{\gamma}
\def\la{\langle}
\def\ra{\rangle}
\newcommand{\ep}{\varepsilon}

\newcommand{\condval}{-(265\pm 5\pm 22 ~ \textrm{MeV})^3}
\begin{document}
\thispagestyle{empty}
\begin{flushright}
\begin{tabular}{l}
{\tt RM3-TH/04-17}\\
{\tt FTUV-04-0908}\\ 
{\tt IFIC/04-49}
\end{tabular}
\end{flushright}

\begin{center}
\vskip 1.1cm
{\LARGE  \bf Operator product expansion and \\
\vskip 0.2cm quark condensate from Lattice QCD\\
\vskip 0.3cm in coordinate space}
\vskip1.3cm 
{\large\sc V.~Gimenez$^{\,a}$, V.~Lubicz$^{\,b,c}$, F.~Mescia$^{\,b,d}$, 
\\ \vskip0.1cm
V.~Porretti$^{\,b,c}$, J.~Reyes$^{\,e}$}\\
\end{center}
\vspace{0.5cm}
 
{\normalsize {\sl
$^a$ Dep. de Fisica Te\`{o}rica and IFIC, Univ. de Val\`{e}ncia, Dr.Moliner 50, E-46100,\\
{\tiny.} \hspace{0.48cm} Burjassot, Val\`{e}ncia, Spain.  \vskip0.3cm
$^b$ Dip. di Fisica, Univ. di Roma Tre,Via della Vasca Navale 84, I-00146 Rome, Italy \vskip0.3cm
$^c$ INFN, Sezione di Roma III,Via della Vasca Navale 84, I-00146 Rome, Italy
\vskip0.3cm
$^d$ INFN, Laboratori Nazionali di Frascati, Via E. Fermi 40, I-00044 Frascati, Italy \vskip0.3cm
$^e$ Dip. di Fisica, Univ. di Roma "La Sapienza", P.le A. Moro 2, I-00185 Rome, Italy }}
\vskip0.5cm
\begin{center}
\end{center}

\begin{abstract} We present a Lattice QCD determination of the chiral quark
condensate based on a new method. We extract the quark condensate from the operator product expansion of the quark propagator at short euclidean  distances, where it represents the leading contribution in the chiral limit. From this study we obtain $\langle\bar q
q\rangle^{\ms}(2\, \textrm{GeV})= \condval$, in good agreement with
determinations of this quantity based on different approaches. The simulation
is performed by using the ${\cal O}(a)$-improved Wilson action at $\beta=6.45$ 
on a volume $32^3\times70$ in the quenched approximation.
\end{abstract}

%\vspace*{0.2cm} 
%{\small{\tt PACS numbers: 11.15.Ha,\ 12.38.Gc,\ 11.10.Gh.}}

\newpage

\setcounter{page}{1}
\setcounter{footnote}{0}
\setcounter{equation}{0}

%%%%%%%%%%%%%%%%%%%%%%%%%%%%%%%%%%%%%%%%%%%%%%%%%%%%%%%%%%%%%%%%%%%%%%%%%%%%%

\section{Introduction}
An accurate determination of the chiral quark condensate is a task of prime
interest. Its non-vanishing value signals the spontaneous breaking of chiral 
symmetry in QCD and, quantitatively, it is related to the pseudo-Goldstone
bosons mass spectrum.

Due to the purely non-perturbative nature of the quark condensate, its estimate
is rather challenging. Traditional approaches have been based on QCD sum rules
(a review of these techniques can be found in refs.~\cite{tech,Dosch:1997wb}). 
In the last years, first principle determinations of the quark condensate have
been provided by Lattice QCD calculations, and the accuracy of these results is
expected to systematically improve in time. The standard method to extract the 
quark condensate from lattice calculations exploits the well known GMOR 
formula~\cite{Giusti:1998wy}-\cite{Hernandez:2001yn}. Alternative techniques
have been also investigated, based on the $\epsilon$-expansion of QCD in a
small volume~\cite{Hernandez:1999cu}-\cite{DeGrand:2001ie} and on the study of
the Goldstone pole contribution to the pseudoscalar quark Green 
function~\cite{Cudell:1998ic,Becirevic:2004qv}.

In this paper, we present an exploratory Lattice QCD determination of the
chiral quark condensate based on a new method. We study the quark
propagator in coordinate space and its operator product expansion
(OPE)~\cite{wilson} at short euclidean distances. The OPE is a powerful
technique that systematically includes  non-perturbative corrections and
parameterizes the non-trivial properties of the  QCD vacuum in terms of
condensates~\cite{cond}. We extract the quark condensate $\langle \bar \psi 
\psi\rangle$ by evaluating the quark propagator at short distances on the 
lattice, and comparing the result with the OPE prediction,
\be
S(x) \;\sim \;C_I(x) \, \frac{\slash x}{(x^2)^2}\; +\; C_m(x)\, \frac{m}{x^2}
\; + \; C_{\bar \psi \psi}(x)\, \langle\bar\psi\psi\rangle\; +\; \ldots
\label{eq:sope}
\ee
where the dots represent higher powers of $x^2$ and of the quark mass 
$m$.\footnote{Throughout this paper we use the notation $x=\sqrt{x^2}$.} Our
final result for the chiral quark condensate, renormalized in the $\ms$ scheme 
at the scale $\mu=2$ GeV, is
\be
\langle\bar q q\rangle^{\ms}(2\, \textrm{GeV})=\condval \,,
\label{eq:final}
\ee
where the first error is statistical and the second systematic. This result is
in good agreement with those obtained from the other methods listed above. It
also provides a remarkable non-perturbative test of the OPE predictions at short
distance in QCD.

The OPE of the quark propagator can be also performed in momentum space, from
which a determination of the quark condensate might be possible as well.  When
working on the lattice with Wilson fermions, however, the leading  contribution
to the OPE in momentum space is a constant term induced by discretization
effects. Though vanishing in the continuum limit, this term is dominant at
fixed lattice spacing with respect to the mass and the condensate
contributions, whose coefficients are suppressed by $1/p^2$ and $1/p^4$
respectively~\cite{Becirevic:1999kb}. In coordinate space this major obstacle
is bypassed, since the Fourier transform of the unphysical term is a
discretized delta function, whose effect is negligible at distances larger than
few lattice  spacings. 

Another advantage of the approach studied in this paper is that it greatly
simplifies the renormalization procedure. Specifically, once the quark
propagator on the l.h.s. of eq.~(\ref{eq:sope}) is renormalized, all
contributions appearing on the r.h.s. turn out to be expressed in terms of
renormalized quantities. In particular, the determination of the chiral quark
condensate in this approach does not require the evaluation of the
corresponding renormalization constant.

The applicability of the OPE to correlation functions evaluated on the lattice 
at fixed value of the lattice spacing $a$ relies on the existence of a short 
distance region where the conditions 
\be
a\simle x \simle 1/\Lam
\label{eq:window}
\ee
are both satisfied. The upper bound in eq.~(\ref{eq:window}) guarantees that 
the Wilson coefficients entering the OPE at the typical scale $\mu=1/x$ can be
evaluated in perturbation  theory. The lower bound must be satisfied in order
to keep under control discretization effects. In the present study, though we
use an ${\cal O}(a)$-improved action and the value of the inverse lattice
spacing is as large as $a^{-1}\simeq$ 4 GeV, we find that in the region $x 
\simle 1/\Lam$ discretization effects in the quark propagator are not
negligible. These effects are in fact responsible for most of the systematic
uncertainty quoted in eq.~(\ref{eq:final}). In order to reduce their 
contribution, we have followed a procedure similar to the one applied in
ref.~\cite{Gimenez:2004me}: we have corrected the lattice results for the
quark propagator by the lattice artifacts computed in the free theory, thus 
reducing their size from ${\cal O}(a^2)$ to ${\cal O}(\as \,a^2)$.

We now summarize the procedure followed in this study and present the plan of 
the paper.

\vspace{0.2truecm}\noindent
-- In sect.2, we derive the OPE of the quark propagator in coordinate space, by
including QCD corrections up to the next-to-leading order (NLO).

\vspace{0.2truecm}\noindent
-- Details of the lattice simulation are presented in sect.3, where the 
tree-level correction of lattice artifacts is also discussed.

\vspace{0.2truecm}\noindent
-- In sect.4 we compute the renormalization constant of the quark 
propagator non-per\-tur\-ba\-ti\-ve\-ly in the $X$-space scheme. The $X$-space 
method has been proposed in ref.~\cite{xspace}, and applied 
in~\cite{Gimenez:2004me} to compute the renormalization constants of bilinear 
quark operators. Our result for the quark field renormalization constant, 
converted to the $\ms$ scheme, reads
\be
  Z_\psi^{\ms}(\mu=2 \gev)= 0.871\pm 0.003\pm 0.020 \,,
\ee
in good agreement with the result obtained in ref.~\cite{Becirevic:2004ny} by
using the non-perturbative RI-MOM method.

\vspace{0.2truecm}\noindent
-- In sect.5 we evaluate the chiral quark condensate by fitting in coordinate
space the quark propagator, extrapolated to the chiral limit, to its OPE. A 
second estimate is obtained by first using the OPE at finite values of the 
quark mass and then extrapolating the result to the chiral limit. Different 
functional forms are considered in the fits, and the differences among the 
results are taken into account in the estimate of the systematic error. The two 
approaches give completely consistent results.

\vspace{0.2truecm}\noindent
-- The final result quoted in eq.~(\ref{eq:final}) is presented in sect.6, 
where we discuss in details the evaluation of the systematic error.

\vspace{0.2truecm}\noindent
-- Finally, we sketch in the appendix the NLO QCD calculation of the Wilson coefficients
entering in eq.~(\ref{eq:sope})

%%%%%%%%%%%%%%%%%%%%%%%%%%%%%%%%%%%%%%%%%%%%%%%%%%%%%%%%%%%%%%%%%%%
%%%%%%%%%%%%%%%%%%%%%%%%%%%%%%%%%%%%%%%%%%%%%%%%%%%%%%%%%%%%%%%%%%%
%%%%%%%%%%%%%%%%%%%%%%%%%%%%%%%%%%%%%%%%%%%%%%%%%%%%%%%%%%%%%%%%%%%
\section{OPE of the quark propagator in coordinate space}
The quark propagator can be expressed in terms of two scalar form factors, 
$\Sigma_1(x)$ and $\Sigma_2(x)$, which are defined from
\be
S(x)= \frac{\slash x}{(x^2)^2}\, \Sigma_1(x)+\frac{1}{x^2}\, \Sigma_2(x).
\label{eq:ciao}
\ee
The leading terms in the OPE of $\Sigma_1(x)$ and $\Sigma_2(x)$ can be read 
from eq.~(\ref{eq:sope}):
\bea
\Sigma_1(x)&=&\frac{1}{2\pi^2}\,C_I(x)+\cdots \nn \\
\Sigma_2(x)&=&\frac{1}{4\pi^2}\,C_m(x)\,m-\frac{1}{4N_c}\,C_{\bar\psi\psi}(x)
\,\langle\bar\psi\psi\rangle \, x^2+\cdots
\label{OPE}
\eea
where, at variance with eq.~(\ref{eq:sope}), the Wilson coefficients $C_I(x),
\,C_m(x)$ and $C_{\bar\psi\psi}(x)$ are normalized to unity in the free 
theory. $N_c$ is the number of colors and the quark condensate is defined as
\be
\langle \bar\psi \psi \rangle \equiv \langle \bar\psi_i^\alpha(0)
\psi_i^\alpha(0)\rangle \,,
\ee
where a summation over repeated color and spin indices is understood. 

By using the known two-loop results for the quark field and the quark mass 
anomalous dimensions in QCD, a simple one-loop calculation provides the 
renormalization group improved expressions for the Wilson coefficients in 
eq.~(\ref{OPE}), at the NLO. The main steps of the calculation are given in the appendix.
We find, in the $\ms$ scheme,
\bea
\Sigma^{\ms}_1(x,\mu)=\dfrac{1}{2\pi^2}\,W_{\psi}(\mu,1/x)
\left[1-2\aspu C_F\xi\left(\g_E-\log 2\right)\right]
\label{PX}
\eea
\bea
\Sigma^{\ms}_2(x,\mu) &=& W_{\psi}(\mu,1/x) \left[1-2\aspu C_F\xi\left(\g_E-
\log 2\right)\right]\nn \\
&& \hspace{-1.2cm}
\left\{ \dfrac{1}{4 \pi^2}W_{m}(\mu,1/x) \left[1+\aspu C_F\left(4-2 (\xi-3)(\g_E-
\log 2)\right)\right] m^\ms(\mu)\right. \label{PI} \\
&& \hspace{-1.2cm}
\left.-\dfrac{1}{4 N_c} W^{-1}_{m}(\mu,1/x)\left[1+2 \aspu C_F\left(1-(\xi+3)(\g_E-
\log 2)\right)\right]\langle\bar\psi\psi\rangle^\ms(\mu)\,x^2 \right\}\,.
\nn
\eea
The terms in square brackets represent the Wilson coefficients at the scale 
$\mu=1/x$, whereas $W_{I}(\mu,1/x)$, with $I=\psi,m$, are the NLO evolution 
functions,
\be
W_{I}(\mu,1/x)=\left(\dfrac{\as(1/x)}{\as(\mu)}\right)^{\frac{\g^0_I}{2\beta_0}}
\left[1+
\left(\dfrac{\beta_1\g^0_I}{2\beta_0^2}-\dfrac{\g^1_I}{2\beta_0}\right)
\dfrac{\as(\mu)-\as(1/x)}{4 \pi}\right]\,.
\label{eq:evol}
\ee
The coefficients of the beta function and of the quark mass and quark field 
anomalous dimensions at the LO and NLO read: 
\bea 
\beta_0 &=& \dfrac{11 N_c-2 n_f}{3} \quad,\quad
\beta_1=\dfrac{34}{3}N_c^2-\dfrac{10}{3} N_c n_f -2 C_F n_f \nn \\
\g^0_m &=& 6\,C_F \quad,\quad 
\g^1_m= C_F \left(\frac{97}{3} N_c+3 C_F-\dfrac{10}{3} n_f \right) \\
\g^0_{\psi} &=& -2 \xi C_F \quad,\quad
\g^1_{\psi}=-4\,C_F \left(\left(\dfrac{25}{8}+\xi+\dfrac{\xi^2}{8}\right) 
N_c-\dfrac{1}{2} n_f- \dfrac{3}{4} C_F\right) \nn
\eea 
where $C_F=(N^2_c-1)/(2 N_c)$, $\xi$ is the gauge parameter ($\xi=0$ in the
Landau gauge) and $n_f$ is the number of active flavors ($n_f=0$ in the 
quenched approximation).

The result in coordinate space for the Wilson coefficient of the quark
condensate given in eq.~(\ref{PI}) corresponds to the one obtained in
ref.~\cite{pasc} in  momentum space. Eqs.~(\ref{PX}) and (\ref{PI}) will be
used in sects.4 and 5 to  extract the quark field renormalization constant and
the chiral quark  condensate with NLO accuracy in the $\ms$ scheme. 

%%%%%%%%%%%%%%%%%%%%%%%%%%%%%%%%%%%%%%%%%%%%%%%%%%%%%%%%%%%%%%%%%%%
%%%%%%%%%%%%%%%%%%%%%%%%%%%%%%%%%%%%%%%%%%%%%%%%%%%%%%%%%%%%%%%%%%%
%%%%%%%%%%%%%%%%%%%%%%%%%%%%%%%%%%%%%%%%%%%%%%%%%%%%%%%%%%%%%%%%%%%
\section{Analysis of discretization effects}
In this section we present the details of the lattice simulation, illustrate the
results obtained for the bare quark propagator and discuss the free theory 
correction implemented in order to reduce the lattice artifacts.

We have generated $180$ gauge configurations in the quenched approximation
with  the non-perturbatively ${\cal O}(a)$-improved Wilson action on a volume 
$32^3\times 70$ at $\beta=6.45$. As a value of the inverse lattice spacing we
use $a^{-1}=3.87(19) \gev$, as obtained in ref.~\cite{masse} from the studies
of  the quark-antiquark potential~\cite{necco} and by using in input the
reference scale $a^{-1}(\beta=6.0)=2.0(1) \gev$.\footnote{Had we used in input
$r_0=0.5\, \textrm{fm}$ we would have obtained $a^{-1}\simeq 4.10\gev$.} We
have computed the quark propagator at four values of the hopping parameter, 
$\kappa=$ 0.1349, 0.1351, 0.1352, 0.1353, corresponding to light quark masses 
in the range $m_s/2 \simle m \simle m_s$. The corresponding values of the
renormalized quark masses have been obtained in ref.~\cite{masse} from the 
study of both the vector and axial-vector Ward identities, and are given in
lattice units, in the $\ms$ scheme, in table~\ref{masse}. These values have 
been used in the study of the OPE of the quark propagator and to perform the 
chiral extrapolations of the quantities we are interested in. The statistical
errors quoted in this paper have been evaluated with the jackknife technique.
%%%%%%%%%%%%%%%%%%%%%%%%%%%%%%%%%%%%%%%%%%%%%%%%%%%%%%%%%%%%%%%%%%%%%%%%%%%%%%
\begin{table}[t]
\begin{center}
\renewcommand{\arraystretch}{1.5}
\begin{tabular}{|c|cccc|} \hline
$\kappa$ & 0.1349 & 0.1351 & 0.1352& 0.1353\\
\hline
$am^{\ms}_{AWI}(2 \gev)$ & 0.0305(4)& 0.0227(3)& 0.0188(2)& 0.0149(2)\\
$am^{\ms}_{VWI}(2 \gev)$ & 0.0288(3)& 0.0215(2)& 0.0178(2)& 0.0141(1)\\
\hline
\end{tabular}
\renewcommand{\arraystretch}{1.0}
\end{center}
\vspace{-0.5cm}
\caption{\it Quark masses in lattice units for the non-perturbatively 
${\cal O}(a)$-improved Wilson action at $\beta=6.45$. The results are taken 
from ref.~\cite{masse} (where they are quoted in the RI-MOM scheme at the scale
$\mu=3 \gev$).} 
\label{masse}
\end{table}
%%%%%%%%%%%%%%%%%%%%%%%%%%%%%%%%%%%%%%%%%%%%%%%%%%%%%%%%%%%%%%%%%%%%%%%%%%%%%%

We have fixed the Landau gauge on the lattice by minimizing the
quantity:
\begin{equation}
\theta=\frac{1}{V}\, \sum_{x}\, {\rm Tr}\, \left[\, \Delta_\mu A_\mu(x)\, \Delta_\nu A_\nu(x)\, \right]
\end{equation}
where $V$ is the lattice volume and $\Delta_\mu A_{\mu}$ is the discretized version of the gauge field divergence $\partial_{\mu}\, A_{\mu}$. We have required
$\theta \leq 5.0 \times 10^{-4}$ for all the configurations used in this study.

\begin{figure}
\renewcommand{\arraystretch}{6.5}
\begin{tabular}{cc}
\epsfxsize7.0cm\epsffile{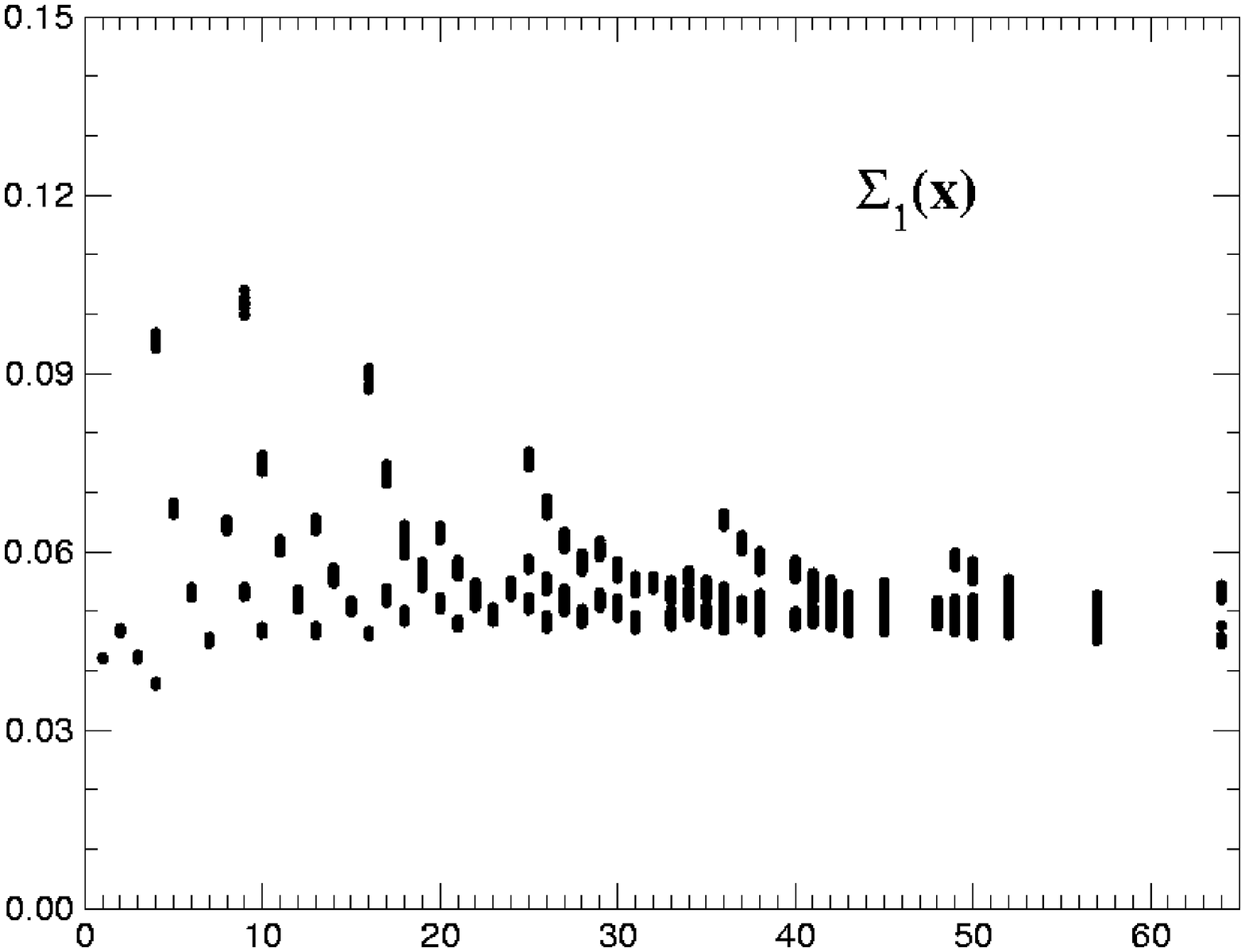} & \hspace*{0.5cm}
\epsfxsize7.0cm\epsffile{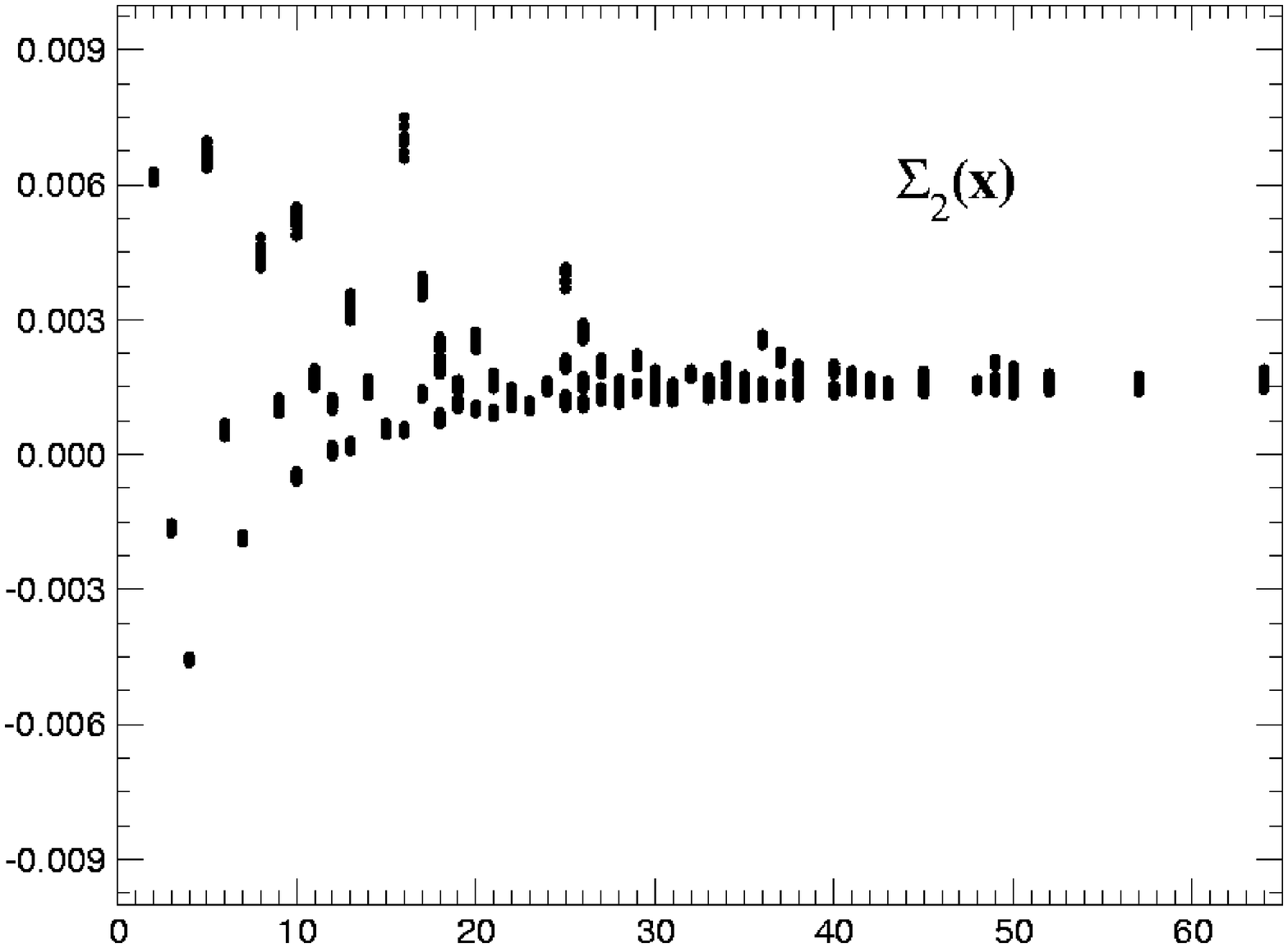} \\
\epsfxsize7.0cm\epsffile{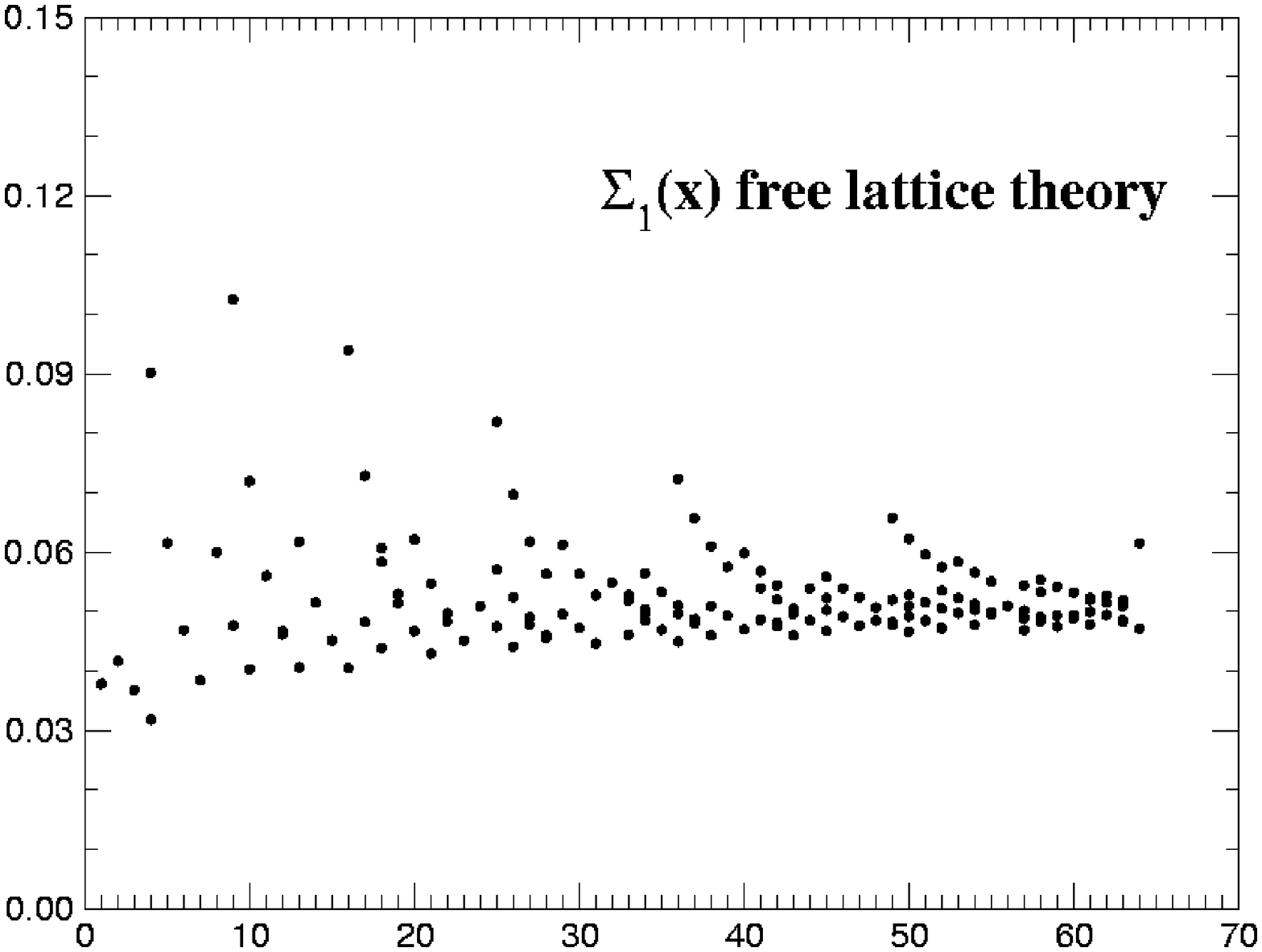} & \hspace*{0.5cm}
\epsfxsize7.0cm\epsffile{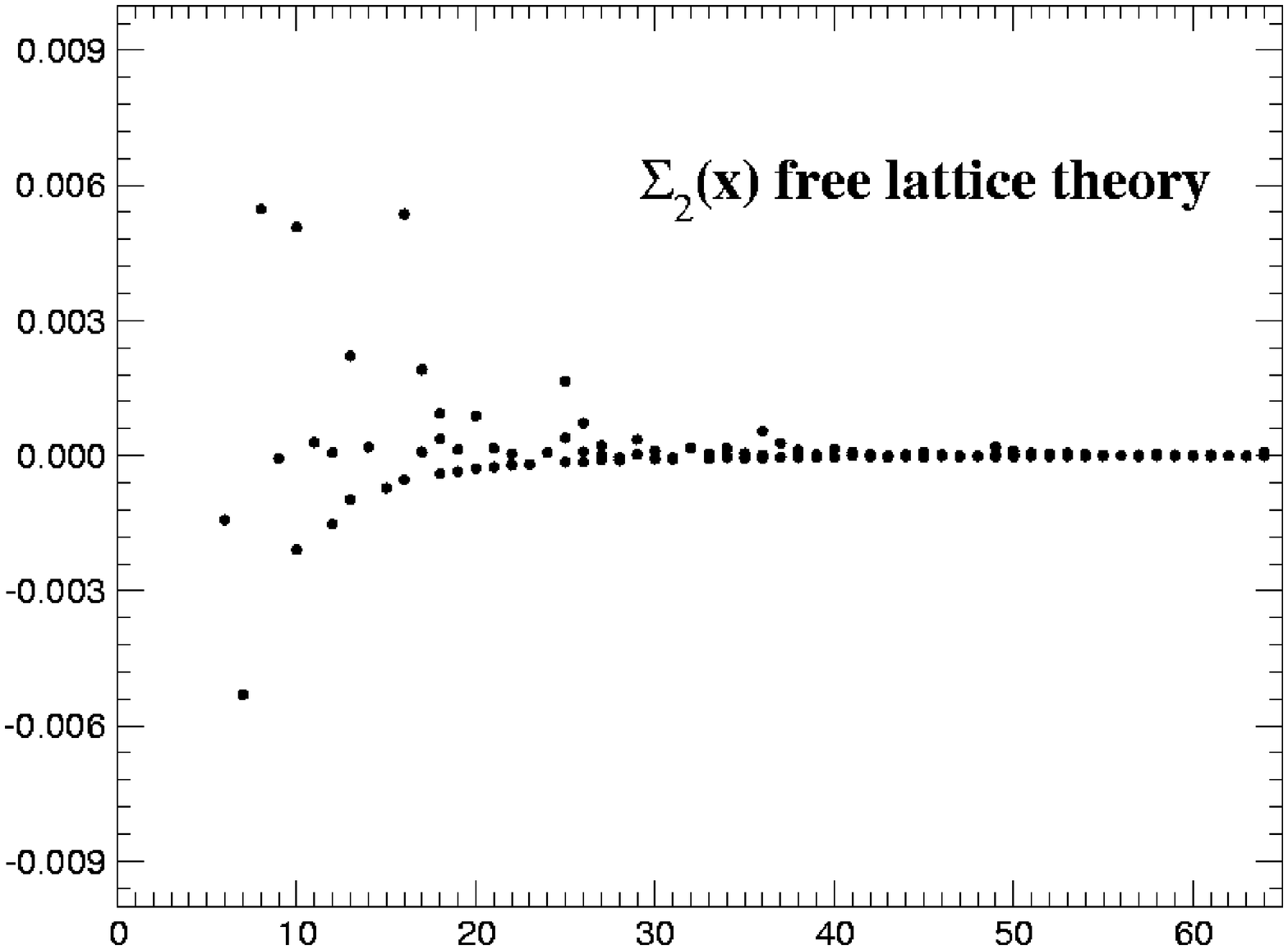} \\
\epsfxsize7.0cm\epsffile{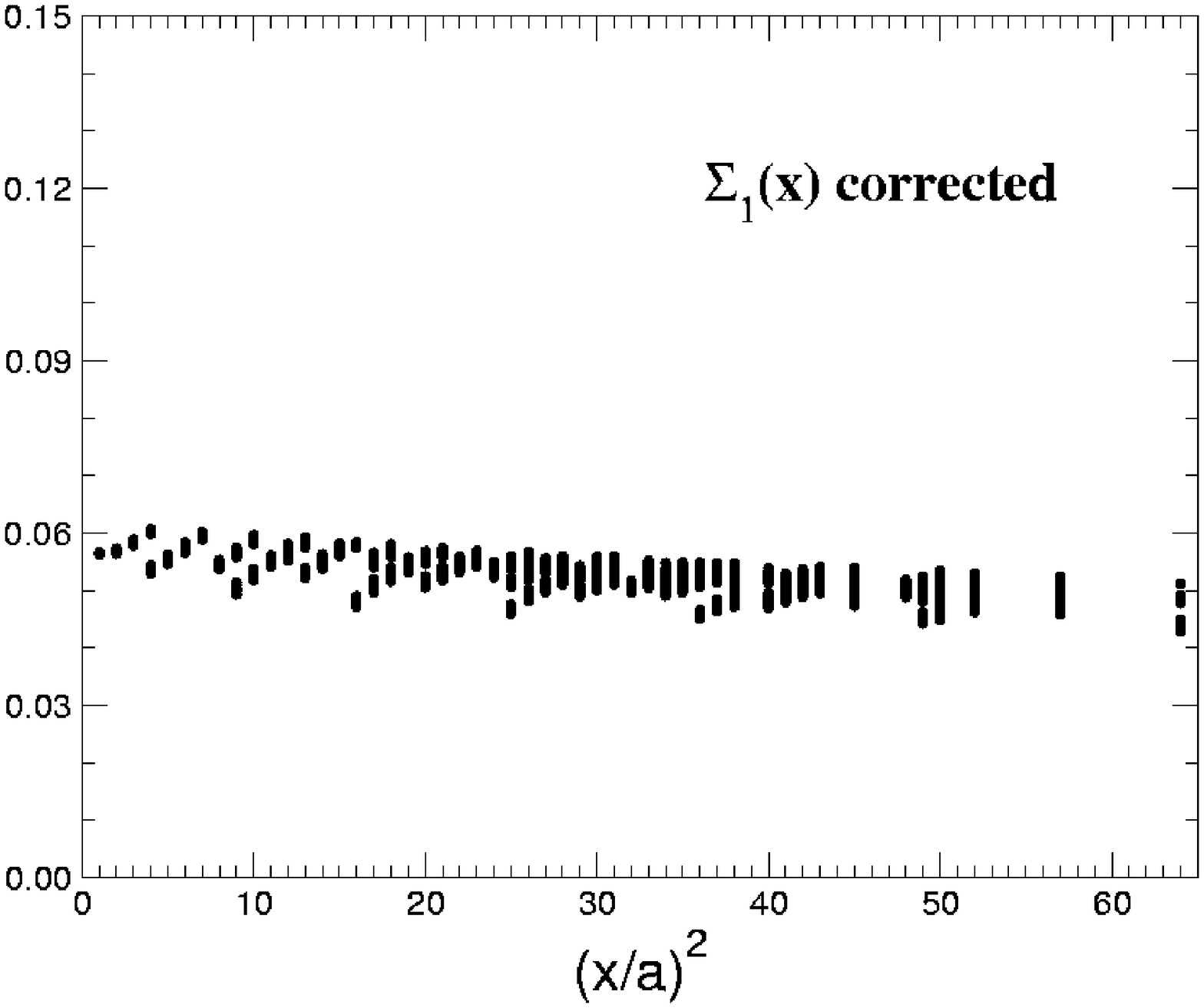} & \hspace*{0.5cm}
\epsfxsize7.0cm\epsffile{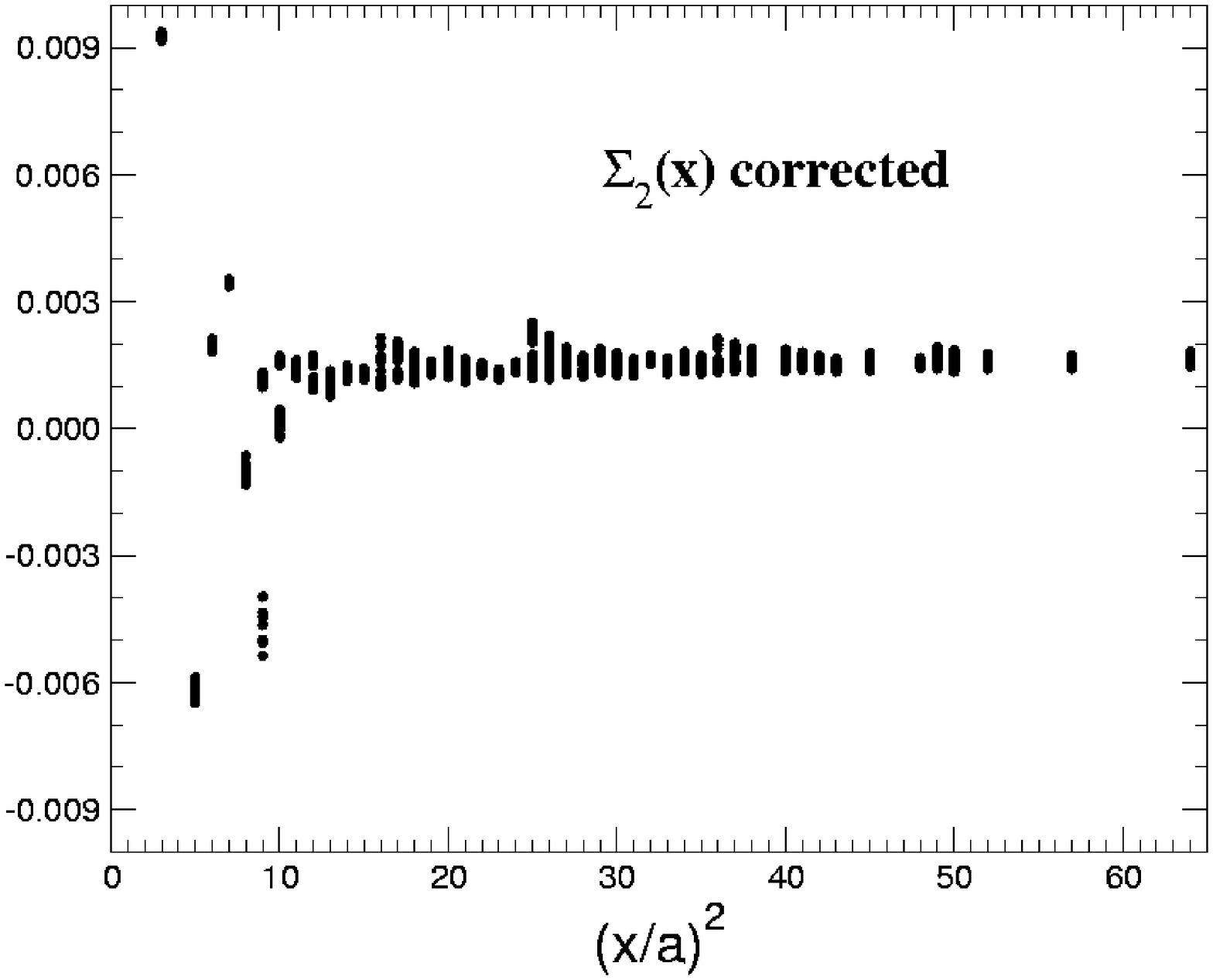}
\end{tabular}
\renewcommand{\arraystretch}{1.0}
\vspace{-0.5cm}
\caption{\it The bare form factors $\Sigma_1$ (left panels) and $\Sigma_2$ 
(right panels). We show, from top to bottom: the form factors in the
interacting theory, as obtained from the lattice simulation at k=0.1349; the
form factors computed in the free lattice theory, at infinite volume and in the
chiral limit; the ``corrected" form factors defined in eqs.~(\ref{eq:s1corr})
and (\ref{eq:s2corr}).}
\label{fig1} 
\end{figure}

In order to compute the quark condensate and the quark field renormalization 
constant, we have extracted from the quark propagator in the Landau gauge
the bare form factors $\Sigma_1(x)$ and $\Sigma_2(x)$ defined in
eq.~(\ref{eq:ciao}). The results are shown in fig.~\ref{fig1}(top) for $\kappa
=0.1349$ as functions of $(x/a)^2$. 
%%%%%%%%%%%%%%%%%%%%%%%%%%%%%%%%%%%%%%%%%%%%%%%%%%%%%%%%%%%%%%%%%%%%%%%%%%%%%
%%%%%%%%%%%%%%%%%%%%%%%%%%%%%%%%%%%%%%%%%%%%%%%%%%%%%%%%%%%%%%%%%%%%%%%%%%%%%

Lorentz invariance requires that, when approaching the continuum limit, the
form factors should become functions of $x^2$ only. At fixed value of the
lattice  spacing, however, the plots in fig.~\ref{fig1}(top) show that points
corresponding to the same value of  $(x/a)^2$ are significantly spreaded out.
This is true especially at short distances ($(x/a) ^2\simle 10$), where
discretization effects are expected to be larger. A better understanding of
these effects can be obtained by studying the lattice quark propagator in the
free theory. Indeed, in the short distance region which is relevant for the
$X$-space method the interacting theory is expected to approach the asymptotic
free regime, up to small perturbative corrections. One finds that $\Sigma_1(x)$
and $\Sigma_2(x)$, computed on the lattice in the free theory, present similar
deviations from the expected continuum behavior. The free theory results,
obtained at infinite volume and in the chiral limit, are shown in
fig.~\ref{fig1}(center), and they can be compared with the lattice results
shown in the top panels. This similarity suggests that one can identify the
discretization patterns in the free case in order to subtract them in the 
interacting case of interest~\cite{Gimenez:2004me}. The practical 
implementation of this approach passes through the definition of the 
``corrected" form factors.

In the case of $\Sigma_1(x)$ we define 
\be
\Sigma_1^{corr}(x)=
\left(\frac{\Sigma^{cont}_{1,\,free}(x)}{\Sigma^{lat}_{1,\,free}(x)}\right) \,
\Sigma_1(x) \,,
\label{eq:s1corr}
\ee
where $\Sigma^{cont}_{1,\,free}(x)$ and $\Sigma^{lat}_{1,\,free}(x)$ are the
free theory form factors computed respectively in the continuum and on the
lattice, at infinite volume and in the chiral limit. For finite values of the
lattice spacing, the difference of the ratio $\Sigma^{cont}_{1,\,free}(x)/
\Sigma^{lat} _{1,\,free} (x)$ from unity is a measure of tree-level 
discretization errors.  After the correction of eq.~(\ref{eq:s1corr}), we 
expect these errors to be reduced from $O(a^2)$ to  $O(\as a^2)$.

Concerning $\Sigma_{2}(x)$ one observes that, in the continuum and in the
chiral limit, the form factor vanishes at any order of perturbation theory.
Therefore, in the case of $\Sigma_{2}$ we implement the following correction: 
\be
\Sigma_2^{corr}(x)=\Sigma_2(x)-\Sigma^{lat}_{2,\,free}(x),
\label{eq:s2corr}
\ee
where $\Sigma^{lat}_{2,\,free}$ represents a pure discretization effect. After
eqs.~(\ref{eq:s1corr}) and (\ref{eq:s2corr}) have been  implemented, we also
average the results for the form factors $\Sigma_1(x)$ and $\Sigma_2(x)$
obtained at lattice points which correspond to the same value of  $x^2$.

The remarkable effect of the correction on the two form factors is shown in 
fig.~\ref{fig1} (bottom). In the following analysis, otherwise indicated, 
we will always use the corrected form factors defined in eqs.~(\ref{eq:s1corr}) 
and (\ref{eq:s2corr}).

%%%%%%%%%%%%%%%%%%%%%%%%%%%%%%%%%%%%%%%%%%%%%%%%%%%%%%%%%%%%%%%%%%%
%%%%%%%%%%%%%%%%%%%%%%%%%%%%%%%%%%%%%%%%%%%%%%%%%%%%%%%%%%%%%%%%%%%
%%%%%%%%%%%%%%%%%%%%%%%%%%%%%%%%%%%%%%%%%%%%%%%%%%%%%%%%%%%%%%%%%%%
\section{Renormalization of the quark propagator in the X-scheme}
In this section, we define the $X$-space renormalization 
scheme~\cite{Gimenez:2004me} for the quark propagator and discuss the
determination of the corresponding renormalization constant. 

The quark field renormalization constant $Z_\psi^X(\mu)$, in the Landau gauge 
$X$-scheme, is determined non-perturbatively by imposing the condition
\be
Z_\psi^X(\mu=1/x)\,\Sigma_1(x)\left\vert^{\xi=0}_{m \to 0}\right. =
\Sigma^{cont}_{1,\,free}(x)
\label{laZ}
\ee
where the value of the form factor in the free continuum theory and in the
chiral limit is $\Sigma^{cont}_{1,\,free}(x)=1/(2\pi^2)$. The limit $m\to 0$ 
in eq.~(\ref{laZ}) guarantees a mass-independent definition of the 
renormalization scheme. It also guarantees that, when using the  ${\cal
O}(a)$-improved Wilson action, the renormalization constant computed from
eq.~(\ref{laZ}) is automatically  ${\cal O}(a)$-improved, without need of
further improving the quark field~\cite{Becirevic:2004ny}.\footnote{In the
definition of the ${\cal O}(a)$-improved quark field, 
\[
q_I(x) = q(x) + a\, b_q\, m\, q(x) + a\, c_q^{\,\prime} (\slash D + m)\, q(x) + 
a\, c_{NGI} \, \slash \partial\, q(x)
\]
the second term vanishes in the chiral limit, while the third one produces in
the quark propagator a contact term in $x=0$. The contribution to the quark
propagator of the last non gauge-invariant term has been found to be
practically indistinguishable from  the contact term proportional to
$c_q^{\,\prime}$~\cite{Becirevic:1999kb}.} 

\begin{figure}[t]
\begin{center}
\includegraphics[width=10cm]{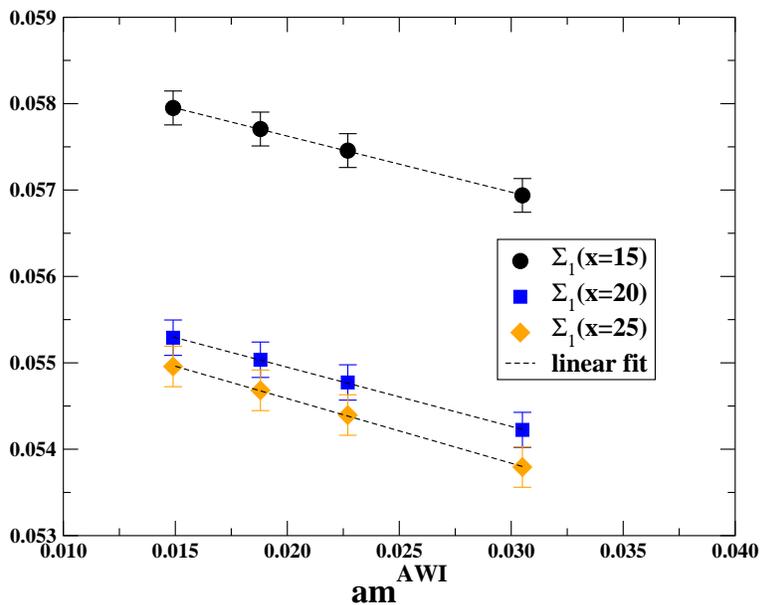}
\end{center}
\vspace{-0.5cm}
\caption{\it Quark mass dependence of $\Sigma_1(x)$ at three different values 
of $x^2$. The dashed lines represent the result of a linear fit.} 
\label{fig:sigma1chiral}
\end{figure}

In order to extrapolate the form factor $\Sigma_1(x)$ to the chiral limit we 
have assumed a linear dependence on the quark mass. This dependence describes 
well the lattice data as can be seen from fig.~\ref{fig:sigma1chiral}, where 
the linear fit is shown for three values of $x^2$ in the range of interest. A 
quadratic fit has been also performed in order to evaluate the systematic error 
involved in the chiral extrapolation.
%%%%%%%%%%%%%%%%%%%%%%%%%%%%%%%%%%%%%%%%%%%%%%%%%%%%%%%%%%%%%%%%%%%%%%%%%%%%%%%
%%%%%%%%%%%%%%%%%%%%%%%%%%%%%%%%%%%%%%%%%%%%%%%%%%%%%%%%%%%%%%%%%%%%%%%%%%%%%%%

By combining the renormalization condition~(\ref{laZ}) with the NLO evolution
function of $Z_\psi$ given in eq.~(\ref{eq:evol}) and considering that the LO
anomalous dimension of the quark field vanishes in the Landau gauge, one finds
at the NLO
\be
Z_\psi^X(\mu)= W_{\psi}(\mu,1/x)\,Z_\psi^X(1/x)=
\left(1-\dfrac{\gamma^1_\psi}{2\beta_0}\dfrac{\as(\mu)-\as(1/x)}{4 \pi}\right) 
\left(\frac{\Sigma^{cont}_{1,\,free}(x)}{\Sigma_1(x)}\right)\,.
\label{pxfit}
\ee
We also note that, in the Landau gauge, the equality of the renormalized form
factor $\Sigma_1(x)$ at one loop in the $\ms$ and $X$ schemes implies that the
NLO anomalous dimensions $\gamma^1_\psi$ are also equal in the two schemes.
In the numerical analysis, $\as(1/x)$ has been evaluated at the NLO in the 
$\ms$ scheme by using $n_f=0$ and the quenched estimate $\Lambda_{QCD}^{n_f=0}
=0.225(21)\gev$ (obtained from $r_0 \Lambda_{QCD}^{n_f=0}=0.602
(48)$~\cite{tedeschi} and using $r_0=0.525(25)$ fm). 

As already discussed in the introduction, the $X$-space non-perturbative
renormalization approach relies on the existence of a window $a\simle x \simle
1/\Lam$ which permits matching the lattice results with the perturbative ones
and, at the same time, to avoid the region at very short distances affected by
contact terms and large  discretization effects. In practice, in the present
study, we consider this  condition satisfied in the range $9\simle (x/a)^2
\simle 25$ (the upper bound  corresponds to $x^{-1}\sim 1 \gev$).

\begin{figure}[t]
\begin{center}
\includegraphics[width=10cm]{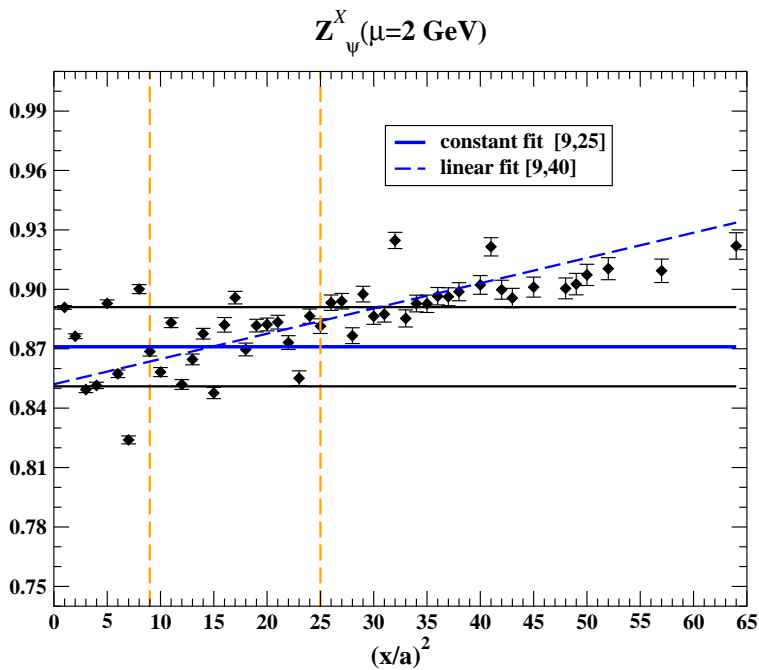}
\end{center}
\vspace{-0.5cm}
\caption{\it Values of $Z_\psi^X(\mu=2 \gev)$ as obtained from 
eq.~(\ref{pxfit}) for different values of $(x/a)^2$. The solid lines indicate 
the results obtained from a constant fit in $x^2$ and the estimated systematic 
error (eq.~(\ref{zpsi})). The dashed vertical lines show the range in $(x/a)^2$ 
where the constant fit is performed. The result of a linear fit in $x^2$ is 
also shown (dashed line). In the latter case, the estimate of $Z_\psi^X(\mu=2
\gev)$ is given by the intercept.}
\label{fig:zpsi}
\end{figure}

The results for $Z_\psi^X(\mu=2\gev)$ as obtained from eq.~(\ref{pxfit}) at 
different values of $x^2$ are shown in fig.~\ref{fig:zpsi}. One observes that 
even in the region $(x/a)^2=[9,25]$ the data show some spread, at the level of
few percent. This spread is due to discretization errors which remain after the
free theory correction has been implemented. It represents the main source of
systematic uncertainty in the evaluation of $Z_\psi$. 
%%%%%%%%%%%%%%%%%%%%%%%%%%%%%%%%%%%%%%%%%%%%%%%%%%%%%%%%%%%%%%%%%%%%%%%%%%%%%%%
%%%%%%%%%%%%%%%%%%%%%%%%%%%%%%%%%%%%%%%%%%%%%%%%%%%%%%%%%%%%%%%%%%%%%%%%%%%%%%%
A second source of uncertainty is due the fact that one cannot exclude, even in
the fitting region $(x/a)^2=[9,25]$, a systematic dependence of the data on
$x^2$, which could be due to higher order contributions to the OPE of
$\Sigma_1(x)$ neglected in eq.~(\ref{OPE}). In order to evaluate this
systematics, we have evaluated $Z_\psi(\mu)$ from both a constant and linear
fit in $(x/a)^2$, by considering in the latter case larger intervals in $x^2$
(up to $(x/a)^2=40$). The results of the fits are presented in
table~\ref{tab:zpsi}.
%%%%%%%%%%%%%%%%%%%%%%%%%%%%%%%%%%%%%%%%%%%%%%%%%%%%%%%%%%%%%%%%%%%%%%%%%%%%%%%
\begin{table}[t]
\begin{center}
\renewcommand{\arraystretch}{1.2}
\begin{tabular}{ccc}
\hline \hline
$Z_\psi^X(\mu)$ &  $(x/a)^2$-range & fit \\
\hline
$0.871\pm 0.003$   &  $[9,25] $ & constant \\
$0.852\pm 0.003$   &  $[9,40] $ &linear \\
%$0.850\pm 0.003$   &  $[15,40]$ &linear \\
\hline 
\hline 
\end{tabular}
\end{center}
\renewcommand{\arraystretch}{1.0}
\vspace{-0.5cm}
\caption{\it $Z_\psi(\mu)$ in the $X$-space scheme at $\mu=2 \gev$ as obtained
from either a constant or a linear fit in $x^2$. The quoted errors are
statistical only.} 
\label{tab:zpsi}
\end{table}
%%%%%%%%%%%%%%%%%%%%%%%%%%%%%%%%%%%%%%%%%%%%%%%%%%%%%%%%%%%%%%%%%%%%%%%%%%%%%%%

The other sources of systematic effects, as those deriving from the
determination of the lattice scale, the estimate of $\Lam$, the difference
between linear and quadratic chiral extrapolations and the use of the vector or
the axial-vector definitions of the quark masses in these extrapolations, are
found to be negligible. As final estimate of  $Z_\psi^{X}(\mu)$ we thus quote
\be
Z_\psi^{X}(2\gev)= 0.871\pm 0.003\pm 0.020
\label{zpsi}
\ee
where the first error is statistical and the second systematic. The central 
value in eq.~(\ref{zpsi}) is the one obtained from the constant fit in the shorter distance range $(x/a)^2=[9,\,25]$, where the contribution of higher power corrections is more suppressed.
The few percent error on the value of $Z_{\psi}$ introduces an uncertainty in the estimate of the chiral quark condensate discussed in the next section which is completely negligible.

The vanishing of the one-loop contribution to the form factor $\Sigma_1(x)$ in 
the Landau gauge implies that the quark field renormalization constant at the
NLO is equal in several commonly used renormalization schemes. In particular,
\be
Z_\psi^X(\mu) = Z_\psi^{\ms}(\mu) = Z_\psi^{\rm{RI-MOM}}(\mu)  
\ee
at the NLO. The result in eq.~(\ref{zpsi}) can be therefore directly compared
to the value $Z_{\psi}^{\rm{RI-MOM}}(2\gev)=0.865 \pm 0.003$ obtained
non-perturbatively in ref.~\cite{Becirevic:2004ny} by using the RI-MOM method. 
It can be also compared with the prediction of one-loop boosted perturbation 
theory $Z_{\psi}^{\ms}(2\gev)\simeq 0.880$. 

We also quote the value of $Z_\psi^{X}$ obtained by using the rough lattice
data, without implementing the tree-level correction of discretization effects: 
$Z^{X}_{\psi}(2\gev)=0.868 \pm 0.003\pm0.080$. The difference in the central
value with respect to eq.~(\ref{zpsi}) is less  than 0.5\%. As expected,
however, the systematic uncertainty is much larger in the latter case, due to
the significantly larger spread of the points in the fitting region. In 
practice, the tree-level correction has smoothed the overall behavior of the 
quark propagator at short distances allowing the reduction of the systematic
uncertainty by about a factor $4$, but affecting the central value by only a 
small amount.

%%%%%%%%%%%%%%%%%%%%%%%%%%%%%%%%%%%%%%%%%%%%%%%%%%%%%%%%%%%%%%%%%%%
%%%%%%%%%%%%%%%%%%%%%%%%%%%%%%%%%%%%%%%%%%%%%%%%%%%%%%%%%%%%%%%%%%%
%%%%%%%%%%%%%%%%%%%%%%%%%%%%%%%%%%%%%%%%%%%%%%%%%%%%%%%%%%%%%%%%%%%
\section{Extraction of the quark condensate}
One of the advantages of the approach considered in this paper to evaluate the
chiral quark condensate is that the renormalization procedure is greatly
simplified: in the OPE of the quark propagator, expressed by 
eq.~(\ref{eq:sope}), once the propagator on the l.h.s. is renormalized by the 
quark field renormalization constant, the r.h.s. turns out to be expressed
directly in terms of renormalized quantities. In particular, the quark 
condensate, renormalized at a scale $\mu$, can be extracted directly from the 
trace of the quark propagator (i.e. the scalar form factor $\Sigma_2$)
renormalized at the same scale. Furthermore, once the quark propagator is 
improved at ${\cal O}(a)$, the operator matrix elements which enter its OPE are
automatically improved at the same order.

In the study of the OPE, the physical quantity which we are interested in is
the quark condensate in the chiral limit. To reach this limit, we have followed
two procedures. In the  first approach, we extrapolate to the chiral limit the
scalar form factor $\Sigma_2(x)$ for each value of $x^2$. The quark condensate
is then evaluated by using the OPE expressed by eq.~(\ref{PI}), which is
accurate at the NLO, in the massless case. In this limit, the quark condensate
represents the leading term of the expansion. In the second approach, which we
consider for a consistency check of the calculation, the order of the
extrapolations is inverted. At finite values of the quark mass, the OPE of
$\Sigma_2$ at order $x^2$ contains, besides the quark condensate, a term
proportional to $m^3$. In this case, we first extract the whole ${\cal O}(x^2)$
contribution to the OPE and then extrapolate the result to the chiral limit. As
we will show in the following, the two procedures yield completely consistent
predictions. We now discuss the two approaches in more detail.
 
\begin{figure}[p]
\begin{center}
\includegraphics[width=10cm]{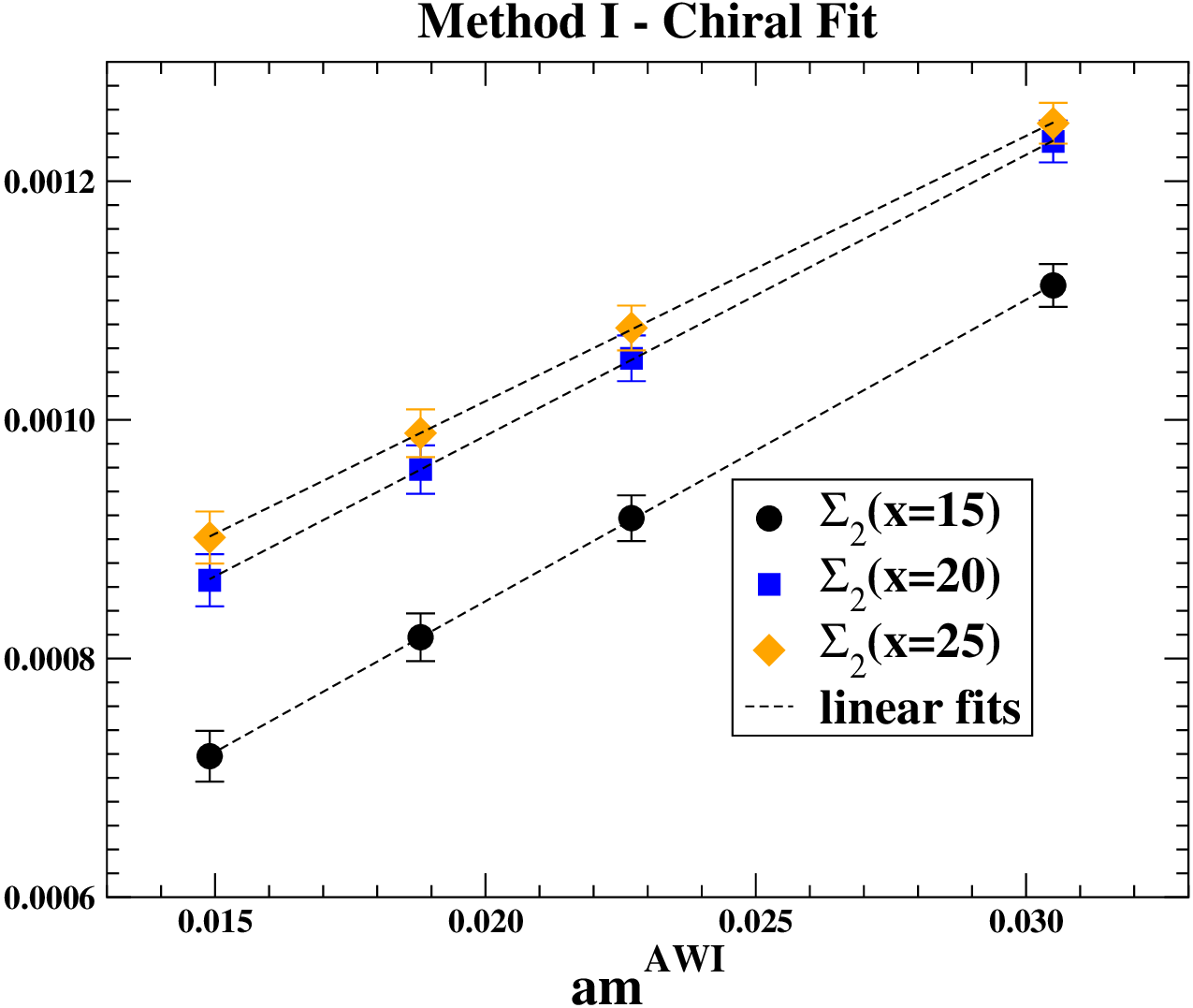}
\end{center}
\vspace{-0.5cm}
\caption{\it Quark mass dependence of $\Sigma_2(x)$ at three different values 
of $x^2$. The dashed lines represent the result of a linear fit.}
\label{fig3}
\end{figure}

\begin{figure}[p]
\vspace{0.5cm}
\begin{center}
\includegraphics[width=10cm]{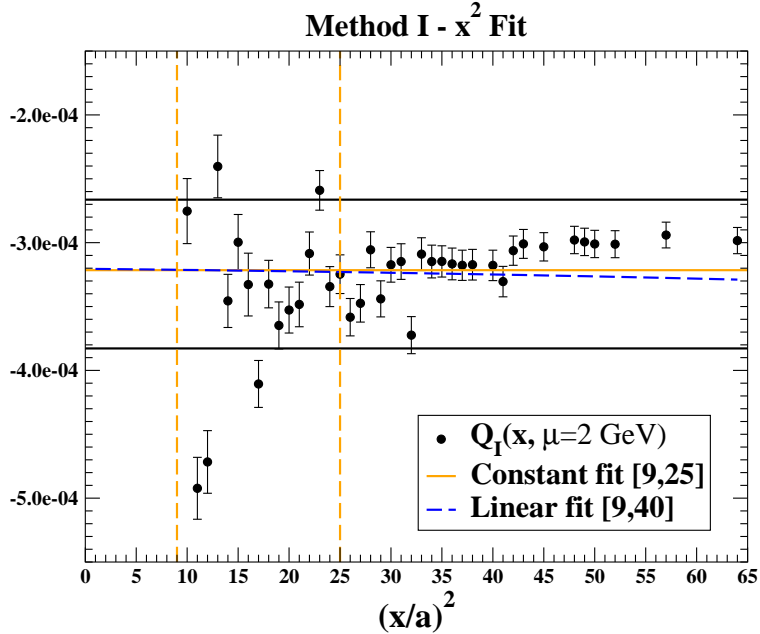}
\end{center}
\vspace{-0.5cm}
\caption{\it Values of $Q_I(x,\mu=2 \gev)$ as a function of $(x/a)^2$.
The solid lines indicate the results obtained from a constant fit in $x^2$ and 
the estimated systematic error. The dashed vertical lines show the range in 
$(x/a)^2$ where the constant fit is performed. The result of the linear fit in 
$x^2$ is also shown (dashed line).} 
\label{fig4}
\end{figure}

\paragraph*{Method I:}
For each value of $x^2$, the renormalized form factor $\Sigma_2(x)$ is 
extrapolated to the chiral limit, both linearly and quadratically in either the 
vector or the axial-vector quark masses.  Examples of this chiral extrapolation,
for three typical values of $x^2$, are shown in fig.~\ref{fig3}. 
For each value of $x^2$ we have then computed the quantity
\be
Q_I(x,\mu) \equiv -\frac{(\Sigma_2^{\ms}(x,\mu))^{\rm chiral}}{C_{\bar\psi\psi}
(x,\mu)\,x^2/4 N_c} = \langle \bar\psi\psi \rangle^\ms(\mu) + {\cal O}(x^2)
\label{eq:qI}
\ee
and performed a fit to the form
\be
Q_I(x,\mu)= \langle \bar\psi\psi \rangle^\ms(\mu) + B\,x^2 \,.
\label{eq:sch}
\ee
Both constant ($B=0$) and linear fits have been performed, and the results are 
presented in table~\ref{tab:condcr}, see also fig.~\ref{fig4}. 
Since the results of the linear fit are unstable when the fit is limited to the
interval $(x/a)^2=[9,\,25]$, we have considered in this case larger  distances,
up to $(x/a)^2=40$. In all cases, we find consistent results for  the quark
condensate, as can be seen from table~\ref{tab:condcr}. We also find that the 
contribution of the ${\cal O}(x^2)$ term is completely negligible, and the 
coefficient B is compatible with zero within the statistical errors. 
%%%%%%%%%%%%%%%%%%%%%%%%%%%%%%%%%%%%%%%%%%%%%%%%%%%%%%%%%%%%%%%%%%%%%%%%%%%%%
\begin{table}[t]
\begin{center}
\begin{tabular}{c|c|c}
\hline \hline
\multicolumn{2}{c|}{$\langle\bar\psi\psi\rangle^{\ms}(\mu=2 \gev)[\mev^{\,3}]$}
& $(x/a)^2$-range \\  
Constant fit & Linear fit & \\ \hline
$-(265\pm 5)^3$   & $-$ & $ [9,25]$\\
$-(266\pm 4)^3$   & $-(265\pm 7)^3$ & $[9,40]$\\
%$-(266\pm 4)^3$   & $-$ & $[15,40]$\\
\hline \hline 
\end{tabular}
\end{center}
\vspace{-0.5cm}
\caption{\it Values of the chiral quark condensate in the $\ms$ scheme at the
scale $\mu=2 \gev$ as obtained from either the constant or the linear fit of
eq.~(\ref{eq:sch}).} 
\label{tab:condcr}
\end{table}
%%%%%%%%%%%%%%%%%%%%%%%%%%%%%%%%%%%%%%%%%%%%%%%%%%%%%%%%%%%%%%%%%%%%%%%%%%%%%
%%%%%%%%%%%%%%%%%%%%%%%%%%%%%%%%%%%%%%%%%%%%%%%%%%%%%%%%%%%%%%%%%%%%%%%%%%%%%%
%%%%%%%%%%%%%%%%%%%%%%%%%%%%%%%%%%%%%%%%%%%%%%%%%%%%%%%%%%%%%%%%%%%%%%%%%%%%%%
%%%%%%%%%%%%%%%%%%%%%%%%%%%%%%%%%%%%%%%%%%%%%%%%%%%%%%%%%%%%%%%%%%%%%%%%%%%%%
%%%%%%%%%%%%%%%%%%%%%%%%%%%%%%%%%%%%%%%%%%%%%%%%%%%%%%%%%%%%%%%%%%%%%%%%%%%%%

As a further check of our results, we have also extracted the quark condensate
directly from the ratio $\Sigma_2/\Sigma_1$ of the two form factors. From
eq.~(\ref{OPE}) one finds that, in the chiral limit, this ratio behaves as
\be
\frac{\Sigma_2^{X}(x,\mu)}{\Sigma_1^{X}(x,\mu)}=\frac{\Sigma_2(x)}
{\Sigma_1(x)}= -\frac{\pi^2}{2 N_c}
\frac{C_{\bar\psi\psi}(x,\mu)}{C_I(x,\mu)} \,
\langle \bar\psi\psi \rangle^\ms(\mu) + {\cal O}(x^2) \,.
\label{eq:rap}
\ee
The determination of the quark condensate from eq.~(\ref{eq:rap}) bypasses the
evaluation of the quark field renormalization constant $Z_\psi$. This constant
cancels in the ratio, since it enters the renormalization of both the form
factors $\Sigma_1$ and $\Sigma_2$. This also implies that the r.h.s. of
eq.~(\ref{eq:rap}) is independent of the choice of the renormalization scale
$\mu$. We also find that this ratio, when computed by using the non-corrected
form factors, exhibits a more stable plateau as a function of $x^2$. The
results for the quark condensate obtained with the two approaches are in
excellent agreement (within less than 2\%), indicating that the uncertainty 
connected with the evaluation of $Z_\psi$ is actually negligible.

\paragraph*{Method II:}
In this second approach we study the OPE of $\Sigma_2(x)$ at finite values of 
the quark mass, extract the ${\cal O}(x^2)$ contribution to the expansion and 
extrapolate it to the chiral limit, in order to get the chiral quark
condensate.

\begin{figure}[p]
\vspace{0.5cm}
\begin{center}
\includegraphics[width=10cm]{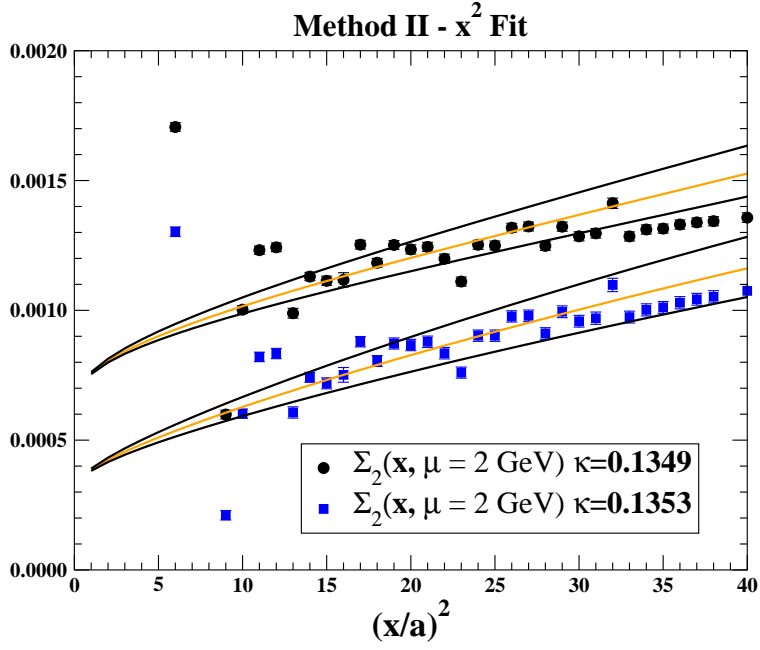}
\end{center}
\vspace{-0.5cm}
\caption{\it Values of $\Sigma_2^{X}(x,\mu=2 \gev)$ as a function of $(x/a)^2$ 
for two different values of the quark mass. The solid lines represent the 
results of the fit to the OPE prediction performed by using for the renormalized
quark masses the values given in table~\ref{masse}.} 
\label{fig5}
\end{figure}

The fit of the form factor $\Sigma_2$ to its OPE is shown in fig.~\ref{fig5}, 
for two values of the quark mass. 
%%%%%%%%%%%%%%%%%%%%%%%%%%%%%%%%%%%%%%%%%%%%%%%%%%%%%%%%%%%%%%%%%%%%%%%%%%%%%%%
%%%%%%%%%%%%%%%%%%%%%%%%%%%%%%%%%%%%%%%%%%%%%%%%%%%%%%%%%%%%%%%%%%%%%%%%%%%%%%%
%%%%%%%%%%%%%%%%%%%%%%%%%%%%%%%%%%%%%%%%%%%%%%%%%%%%%%%%%%%%%%%%%%%%%%%%%%%%%
\begin{figure}[p]
\vspace{1.0cm}
\begin{center}
\includegraphics[width=10cm]{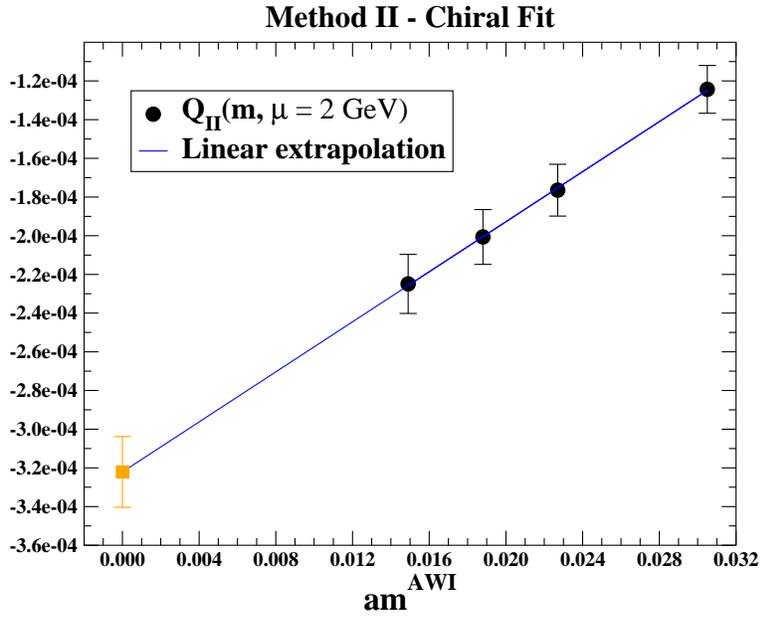}
\end{center}
\vspace{-0.5cm}
\caption{\it Linear fit of $Q_{II}(m,\mu=2 \gev)$ as a function of the quark 
mass. The extrapolated value is the chiral quark condensate in lattice units.}
\label{ultima}
\end{figure}
%%%%%%%%%%%%%%%%%%%%%%%%%%%%%%%%%%%%%%%%%%%%%%%%%%%%%%%%%%%%%%%%%%%%%%%%%%%%%
We find that the mass term contribution to the OPE, which is leading at very
short distances where lattice artifacts are more severe, is poorly estimated 
from the fit. For this reason, we have chosen to fix the renormalized quark 
mass in eq.~(\ref{PI}) to the values determined in ref.~\cite{masse} and 
collected in table~\ref{masse}. Therefore, for each value of the quark mass we 
compute the quantity
\be
Q_{II}(x,m,\mu) \equiv -\frac{\Sigma_2^{\ms}(x,\mu)-C_{m}(x,\mu)\, m^\ms(\mu)
/4 \pi^2}{C_{\bar\psi\psi}(x,\mu)\,x^2/4 N_c} \,.
\label{eq:qII}
\ee
Notice that, in the chiral limit, $Q_{II}(x,m,\mu)$ reduces to $Q_{I}(x,\mu)$ 
defined in eq.~(\ref{eq:qI}). The small spread of the results coming from 
choosing the vector or the axial-vector quark masses is included in the 
systematics. We then fit $Q_{II} (x,m,\mu)$ either to a constant or linearly in
$x^2$ and extrapolate the result, denoted as $Q_{II}(m,\mu)$ in 
fig.~\ref{ultima}, to the chiral limit, where it reduces to the chiral quark
condensate. The quark mass extrapolation is shown in fig.~\ref{ultima}. Though
the points in the plot look very well aligned, a quadratic fit in the  quark
mass has been also performed, in order to evaluate the corresponding systematic
uncertainty. We find that the results obtained for the chiral quark  condensate
with this second approach are indistinguishable, within the statistical errors,
from those derived by using the method I and presented in 
tab.~\ref{tab:condcr}.

%%%%%%%%%%%%%%%%%%%%%%%%%%%%%%%%%%%%%%%%%%%%%%%%%%%%%%%%%%%%%%%%%%%
%%%%%%%%%%%%%%%%%%%%%%%%%%%%%%%%%%%%%%%%%%%%%%%%%%%%%%%%%%%%%%%%%%%
%%%%%%%%%%%%%%%%%%%%%%%%%%%%%%%%%%%%%%%%%%%%%%%%%%%%%%%%%%%%%%%%%%%
\section{Results and discussion}
Our final estimate for the chiral quark condensate is obtained from the results 
given in table~\ref{tab:condcr} after including the evaluation of the 
systematic error. We quote: 
\begin{equation}
\langle\bar \psi \psi \rangle^{\ms}(2 \textrm{GeV})=\condval \,,
\end{equation}
where the first error is statistical and the second systematic. The latter, 
which amounts to about 25\%, is due to:
\begin{itemize}
\item[-] the spread of the points in the fitting regions. As discussed in 
sect.3, this spread is mostly due to discretization effects which are left
after the tree-level ${\cal O}(a)$-correction has been applied to the lattice
data. This error, of about 18\%, represents the main source of systematic
uncertainty, besides the quenching approximation.
\item[-] Arbitrariness in the choice of the fitting interval (within the window 
$a\simle x\simle 1/\Lam$). This yields a 4-5\% uncertainty.
\item[-] Different functional forms considered in the fits. Performing a 
quadratic fit instead of a linear one in the quark mass extrapolations 
introduces a systematic difference of about 9\%. Including the ${\cal O}(x^2)$ 
contribution in the fits of $Q_I$ and $Q_{II}$ to their OPEs gives a 4-5\% 
variation in the results.
\item[-] The statistical error associated to the determination of the lattice 
spacing. This error introduces an uncertainty of about 15\% in the estimate of
the quark condensate. Notice that the systematic error associated in the 
quenched approximation to the dependence of the lattice spacing on the physical 
quantity used to fix the scale is not included. We consider this error as a 
part of the systematic quenching effect. 
\item[-] The uncertainty on the quark field renormalization constant $Z_\psi$,
used to renormalize the quark propagator. This effect is completely negligible 
in the determination of the quark condensate, as discussed in sect.5. 
\item[-] The difference between the results obtained using either the vector or
the axial-vector definitions of the quark masses. The systematics is slightly 
affected by this effect, by less than 1\%.
\item[-] The 10\% error on the quenched estimate of $\Lam$ gives a completely
negligible uncertainty in the determination of the quark condensate.
\end{itemize}

The uncertainty coming from finite volume effects cannot be directly estimated
in the present study, since our results have been obtained at fixed volume. A
study of lattice artifacts performed in ref.~\cite{Gimenez:2004me} has shown
that in the short distance region, which is the one relevant for the $X$-space
method, finite volume effects on the lattice correlation functions in the free
theory are negligible with respect to discretization effects. We expect this
result to remain valid in the interacting theory as well, though a more
quantitative conclusion on this point would require further investigations. The
main source of uncertainty which is not evaluated in our estimate of the chiral
quark condensate is the effect of the quenching approximation.

In conclusion, in this exploratory study we have investigated on the lattice 
the OPE of the quark propagator at short euclidean distances, and shown the 
feasibility of this approach to compute the chiral quark condensate. The result
obtained in this way is in good agreement with previous determinations of this
quantity based on different approaches. The strategy investigated in the
present study can be also applied to compute on the lattice the matrix elements
of other local operators which enter the OPE of correlation functions at the 
leading orders. It can be also directly implemented in lattice simulations
performed with dynamical quarks.

\section*{Acknowledgments}
We thank D.~Becirevic, L.~Giusti, G.~Martinelli and M.~Testa for useful 
discussions and comments on the manuscript. The  work of  F.M.~is partially
supported by IHP-RTN, EC contract  No.~HPRN-CT-2002-00311 (EURIDICE), the work
of V.G. by MCyT, Plan National  I+D+I (Spain) under the grant BFM2002-00568.

%%%%%%%%%%%%%%%%%%%%%%%%%%%%%%%%%%%%%%%%%%%%%%%%%%%%%%%%%%%%
%%%%%%%%%%%%%%%%%%%%%%%%%%%%%%%%%%%%%%%%%%%%%%%%%%%%%%%%%%%%
\section*{Appendix: NLO calculation of the Wilson coefficients}

In this appendix we sketch the NLO QCD calculation of the Wilson coefficients introduced in eq.~(\ref{OPE}).
 
The OPE of the quark propagator in euclidean space is expressed by 
\bea
T(\psi(x)\bar\psi(0))&=& 
\dfrac{1}{2\pi^2}\, C_I(x)\, \dfrac{(\slash x)}{(x^2)^2}+ 
\dfrac{1}{4\pi^2}\, C_m(x)\, \dfrac{m}{x^2} \nn\\
&-&\dfrac{1}{4 N}\, C_{\bar\psi\psi}(x) \, (\bar\psi \psi) + \ldots 
\label{eq:macth}
\eea
where the dots represent higher powers of $x^2$ and of the quark mass $m$. All quantities in eq.~(\ref{eq:macth}) are renormalized at a given scale and in a given renormalization scheme. In the following, we will choose the $\ms$ renormalization scheme.

In order to determine the Wilson coefficients at the NLO in QCD, we calculate  
both the left and the right hand side of eq.~(\ref{eq:macth}) up to ${\cal O}(\as)$ by choosing a common set of external states. The coefficients $C_I(x)$ and $C_m(x)$, in particular, can be determined by taking the vacuum expectation value of eq.~(\ref{eq:macth}) in perturbation theory, where the contribution of the quark condensate is vanishing. Eq.~(\ref{eq:macth}) then simply reduces to
\be
S(x)=\dfrac{1}{2\pi^2}\, C_I(x)\, \dfrac{(\slash x)}{(x^2)^2} + 
\dfrac{1}{4\pi^2}\, C_m(x)\, \dfrac{m}{x^2} \,,
\label{eq:sx}
\ee  
where $S(x)$ is the quark propagator computed in one-loop perturbation theory. By using dimensional regularization, with $D=4-2\ep$, one has
\be
S(x)=\dint \dfrac{d^D\, k}{(2\pi)^{D}} \, e^{-i k\cdot x}S(k)
\label{eq:m1}
\ee
where 
\bea
S(k) &=& Z_{\psi}\, \frac{\slash k}{i\,k^2}\left[1 - \frac{\as}{4 \pi} \,C_F\, 
\xi \, \left(\dfrac{k^2}{\mu^2}\right)^{-\ep} \left(\frac{1}{\hat\ep} + 1\right) \right] + \nn \\
&+& Z_{\psi}\, \frac{Z_m^{-1} \,m}{k^2} \left[1+\frac{\as}{4 \pi} \, C_F  \, \left(\dfrac{k^2}{\mu^2}\right)^{-\ep} \left(\frac{3-\xi}{\hat\ep} +4 \right)
\right]\,,
\label{eq:sk}
\eea
and $1/\hat\ep \equiv 1/\ep + \log (4\pi)-\g_E$. From eq.~(\ref{eq:sk}) one derives the expressions of the quark field and the quark mass renormalization constants in the $\ms$ scheme:
\be
Z_{\psi} = 1+\dfrac{\as}{4 \pi} \, C_F \, \frac{\xi}{\hat\ep} \quad,\quad
Z_m = 1+\dfrac{\as}{4 \pi} \, C_F \, \frac{3}{\hat\ep} \,.
\ee
By inserting eq.~(\ref{eq:sk}) into eq.~(\ref{eq:m1}) and using
\be
\dint \dfrac{d^D\, k}{(2\pi)^{D}} \frac{e^{-i k\cdot x}}{(k^2)^r}=
\dfrac{1}{(4\pi)^{D/2}} \, \dfrac{\Gamma(D/2-r)}{\Gamma(r)} \left( \frac{x^2}{4}\right)^{r-D/2}
\label{eq:fourier}
\ee
we obtain
\bea
S(x)&=&
\dfrac{1}{2\pi^2} \left[1-\dfrac{\as}{4 \pi} \, C_F \, \xi \left(2 
\g_E + \log(\mu^2 x^2/4)\right) \right] 
\dfrac{(\slash x)}{(x^2)^2}+ \nn\\
&&\dfrac{1}{4\pi^2}\left[1+\dfrac{\as}{4 \pi}\, C_F\, \left(4- (\xi-3)\left(2 \g_E + \log(\mu^2 x^2/4) \right) \right)\right]
\dfrac{m}{x^2} \,.
\label{coeffi1}
\eea
From this result, after comparing with eq.~(\ref{eq:sx}), the Wilson coefficients $C_I(x)$ and $C_m(x)$ can be readily identified:
\bea
C_I(x) &=& 1-\dfrac{\as}{4 \pi} \, C_F \, \xi \left[2 
\g_E + \log(\mu^2 x^2/4)\right] \nn \\
C_m(x) &=& 1+\dfrac{\as}{4 \pi}\, C_F\, \left[4- (\xi-3)\left(2 \g_E + \log(\mu^2 x^2/4) \right) \right] \,.
\eea

In order to compute the Wilson coefficient of the quark condensate, $C_{\bar\psi\psi}(x)$, we derive a matching equation by inserting  both sides of eq.~(\ref{eq:macth}) in a connected Green function with two external quark fields at fixed momenta: 
\bea
\la (\psi_\alpha^a(x)\bar\psi_\beta^b(0))
\bar\psi_\gamma^c(p_1) \psi_\delta^d(p_2) \ra =- \frac{1}{4N} \,
\delta_{\alpha\beta} \delta^{ab} C_{\bar\psi\psi}(x)\, \la (\bar\psi(0) \psi(0)) \bar\psi_\gamma^c(p_1) \psi_\delta^d(p_2) \ra \textrm{.}
\label{eq:m3}
\eea
By putting
\be
C_{\bar\psi\psi}(x)= C^0_{\bar\psi\psi}(x)+\frac{\as}{4\pi} \, C^1_{\bar\psi\psi}(x)
\ee
and summing in eq.~(\ref{eq:m3}) over Dirac ($\alpha$,$\beta$) and color ($a$,$b$) indices, one immediately obtains the ${\cal O}(1)$ contribution to the Wilson coefficient,
\be
C^0_{\bar\psi\psi}=1 \,.
\label{eq:c0}
\ee

\begin{figure}[t]
\begin{center}
\includegraphics[angle=270,width=7.0cm]{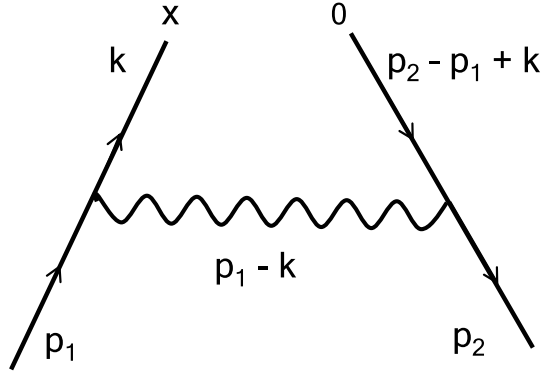}
\end{center}
\caption{\it Feynman diagram relevant for the calculation of the Wilson coefficient $C_{\bar\psi\psi}(x)$ at ${\cal O}(\as)$.} 
\label{app}
\end{figure}

At ${\cal O}(\as)$, the matching equation can be schematically written as
\be
\la (\bar\psi(0)\psi(x)) \bar\psi\psi\ra_1=C^0_{\bar\psi\psi}\la (\bar\psi(0)\psi(0)) \bar\psi\psi\ra_1 + C^1_{\bar\psi\psi} \la (\bar\psi(0) \psi(0)) \bar\psi\psi\ra_0
\label{eq:m2}
\ee
where $\la \ldots \ra_0$ and $\la \ldots \ra_1$ represent respectively the ${\cal O}(1)$ and ${\cal O}(\as)$ contributions to the Green functions.  We now consider the amputated Green functions, and use eq.~(\ref{eq:c0}) together with
the relation
\be
\la (\bar\psi(0)\psi(0)) \bar\psi\psi\ra_0^{amp}=I \,,
\ee 
to obtain
\be
C^1_{\bar\psi\psi}\cdot I=\la (\bar\psi(0)\psi(x)) \bar\psi\psi\ra_1^{amp}- \la (\bar\psi(0)\psi(0)) \bar\psi\psi\ra_1^{amp} \,.
\label{eq:c1}
\ee
The Feynman diagram which contributes to eq.~(\ref{eq:c1}) at ${\cal O}(\as)$ is shown in fig.~\ref{app}. 
%%%%%%%%%%%%%%%%%%%%%%%%%%%%%%%%%%%%%%%%%%%%%%%%%%%%%%%%%%%%
%%%%%%%%%%%%%%%%%%%%%%%%%%%%%%%%%%%%%%%%%%%%%%%%%%%%%%%%%%%%
Since the matching condition is independent of the choice of the external states, we can evaluate this diagram by putting directly $p_1=p_2=0$. In addition, by having neglected in eq.~(\ref{eq:macth}) higher power corrections in the quark mass, we can compute the amputated Green functions in eq.~(\ref{eq:c1}) directly in the limit $m=0$. We then find:
\bea
C^1_{\bar\psi\psi} \cdot I &=&16\, \pi^2 C_F \, \mu^{2 \ep}\dint \dfrac{d^D\, k}{(2\pi)^{D}} \,\left( e^{-i k\cdot x}-1\right) \, \left[\g^\mu \frac{1}{\slash k} \frac{1}{\slash k}\g^\nu \right]\,\frac{1}{k^2}
\left(\delta_{\mu\nu}-(1-\xi)\frac{k_\mu k_\nu}{k^2}\right) + \nn \\
&& \left[\left(Z_\psi -1 \right) - \left(Z_{\bar \psi \psi} -1 \right)\right] \cdot I
\label{eq:d0fey2}
\eea
where $ Z_{\bar \psi \psi} = Z_m^{-1}$. By evaluating the Feynman integral with the aid of eq.~(\ref{eq:fourier}), we finally obtain
\be
C_{\bar\psi\psi}(x) =  1 + \frac{\as}{4\pi}\, C_F \left[ 2 - (\xi+3)\left( 2\g_E + \log(\mu^2 x^2/4)\right)\right]
\label{coeffi2}
\ee
The complete NLO expressions for the Wilson coefficients are derived from eqs.~(\ref{coeffi1}) and (\ref{coeffi2}) by applying the standard NLO evolution functions introduced in eq.~(\ref{eq:evol}).

%%%%%%%%%%%%%%%%%%%%%%%%%%%%%%%%%%%%%%%%%%%%%%%%%%%%%%%%%%%%%%%%%%%%%%%%%%%%%%
%%%%%%%%%%%%%%%%%%%%%%%%%%%%%%%%%%%%%%%%%%%%%%%%%%%%%%%%%%%%%%%%%%%%%%%%%%%%%%

\end{document}